\documentclass[
aip,
amsmath,amssymb,
reprint,
]{revtex4-2}

\usepackage{upgreek}
\usepackage{graphicx}
\usepackage{dcolumn}
\usepackage{bm}
\usepackage{mathptmx}
 \usepackage{subdepth}
\usepackage{hyperref}
\hypersetup{
     colorlinks   = true,
     citecolor    = blue,
     urlcolor=blue,
     linkcolor=blue
}

\usepackage{tabularx}
\usepackage{xcolor}
\usepackage{siunitx}
\usepackage{csquotes}
\usepackage{chemformula}

\newcommand{\tMn}{$\uptau$-MnAl}

\begin{document}

\title{Anomalous Nernst effect in perpendicularly magnetised $\uptau$-MnAl thin films}

\author{D. Scheffler}
\affiliation{Institute for Solid State and Materials Physics, Technische Universität Dresden, 01062 Dresden, Germany}

\author{S.Beckert}%
\affiliation{Institute for Solid State and Materials Physics, Technische Universität Dresden, 01062 Dresden, Germany}

\author{H.Reichlova}%
\affiliation{Institute for Solid State and Materials Physics, Technische Universität Dresden, 01062 Dresden, Germany}
\affiliation{Institute of Physics ASCR, v.v.i., Cukrovarnick\'a 10, 162 53, Praha 6, Czech Republic}

\author{T.G. Woodcock}
\affiliation{Leibniz Institute for Solid State and Materials Research Dresden (IFW Dresden), 01069 Dresden, Germany}

\author{S.T.B. Goennenwein}
\affiliation{Department of Physics, University of Konstanz, 78457 Konstanz, Germany}

\author{A. Thomas}%
\affiliation{Institute for Solid State and Materials Physics, Technische Universität Dresden, 01062 Dresden, Germany}
\affiliation{Leibniz Institute for Solid State and Materials Research Dresden (IFW Dresden), 01069 Dresden, Germany}

\date{\today}

\begin{abstract}

\tMn~is interesting for spintronic applications as a ferromagnet with perpendicular magnetic anisotropy due to its high uniaxial magnetocrystalline anisotropy. Here we report on the anomalous Nernst effect of sputter deposited \tMn~thin films. We demonstrate a robust anomalous Nernst effect at temperatures of 200\,K and 300\,K with a hysteresis similar to the anomalous Hall effect and the magnetisation of the material. The anomalous Nernst coefficient of $(0.6\pm 0.24)\,\upmu$V/K at 300\,K is comparable to other perpendicular magnetic anisotropy thin films. Therefore \tMn~is a promising candidate for spin-caloritronic research.

\end{abstract}
\maketitle
\section{\label{sec:Intro}Introduction}

\tMn~is a L1$_0$ ordered ferromagnetic compound with a high Curie temperature of approx. 650\,K, a saturation magnetisation of 600 kAm$^{-1}$ and a high uniaxial magnetic anisotropy of 1.7 MJm$^{-3}$ making it an interesting material for the application as a rare earth free permanent magnet \cite{Coey2014,Kontos2020,Feng2020}. The material also displays perpendicular magnetic anisotropy (PMA), which is crucial for many spintronic applications, e.g.\ magnetic storage or spin transfer torque switching \cite{Apalkov2016,Ikeda2010,Cubukcu2014,Avci2012}. Consequently, the magnetotransport properties of \tMn~thin films were studied in terms of the anomalous Hall effect (AHE) and tunnel magnetoresistance \cite{Takeuchi2022,Zhang2017,Zhu2016a,Zhu2016b,Saruyama2013,Meng2017a,Meng2017b,Luo2016,Anuniwat2013}. However, the magneto-thermal properties of \tMn~are still unknown, despite renewed interest in large anomalous Nernst (ANE) effect values in emerging materials \cite{Asaba2021,Sakai2018,He2021}, which are often discussed in the context of intrinsic Berry phase related contributions \cite{Shi2020}. The large ANE values also renewed interest in its application potential as, e.g., a heat flux sensor \cite{Zhou2020} or energy harvesting elements \cite{Mizuguchi2019}, in particular, in PMA materials \cite{Hasegawa2015,Ando2016,Xu2019}.

In this manuscript, we report on the anomalous Nernst effect (ANE) in epitaxial thin films of \tMn~grown via magnetron sputtering at various temperatures. To measure the ANE (the thermal counterpart of the anomalous Hall effect), a thermal gradient $\nabla T_{x}$ applied to the sample drives a voltage perpendicular to both the thermal gradient $\nabla T_{x}$ and the magnetic order vector. We show that the ANE is spontaneous (present at zero magnetic field) and that it exhibits a robust hysteresis with a coercivity of approximately $\mu_0H_\mathrm{c}\approx1$\,T, comparable the $H_\mathrm{c}$ values obtained from magnetometry measurements with a superconducting quantum interference device vibrating sample magnetometer (SQUID-VSM) and the AHE. We compare the anomalous Nernst coefficient $S_{xy}^{\mathrm{A}}$ to other PMA materials and find  that \tMn~thin films are an interesting candidate material for spintronic and spin-caloritronic research and devices.
\phantom{}

\phantom{}

\phantom{}

\phantom{}

\section{\label{sec:Experimental Methods}Film growth, structure and magnetic properties}

MnAl thin films were deposited by magnetron sputtering onto MgO(001) substrates using a Bestec UHV sputter tool. The films investigated in this manuscript had a thickness of $t_{\ch{MnAl}}=\SI{78}{nm}$ and a Cr(001) buffer layer ($t_{\mathrm{Cr}}=19$\,nm) was utilised to reduce the lattice mismatch between the \ch{MgO} substrate and the thin films \cite{Oshima2020,Hosoda2012,Saruyama2013}. 

Prior to deposition, the substrates were subsequently cleaned in acetone and ethanol in a ultrasonic bath for 10 minutes (each). After transferring to the chamber, the substrates were annealed at 800°C for 1\,h to eliminate surface contaminations. The Cr buffer layer was deposited at room temperature with a power of 30\,W (DC) using a sputtering pressure of $p_{\mathrm{dep,Cr}}=3\times10^{-3}$mbar (Ar). A subsequent annealing step at 600°C for 1\,h in vacuum was necessary for the crystallisation of Cr. The MnAl film was deposited by cosputtering, applying 60\,W rf power to a MnAl(50/50 at.\%) target and 10\,W dc power to an aluminium target under a sputtering pressure of $p_{\mathrm{dep,Al}}=5\times10^{-3}$mbar (Ar). A temperature of 300\,°C during the deposition and an in situ post-annealing at 500\,°C for 1\,h under vacuum is needed for the crystallisation and the chemical ordering of \tMn. While several samples were fabricated and measured, in this work we focus on the results of one representative \tMn~thin film. Similar results were ahieved for the other \tMn~thin films.

\begin{figure}[ht]
\includegraphics{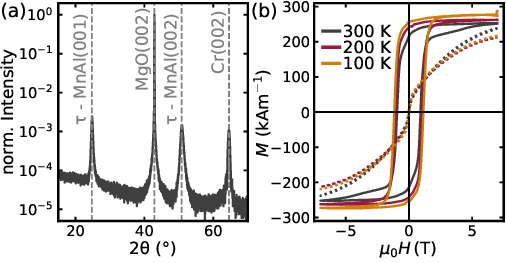}
\caption{\label{fig:SQUID XRD}(a) Symmetrical 2$\theta-\omega$ XRD scan of 78 nm thick \tMn~thin film. (b) SQUID-VSM measurement of the same thin film at different temperatures. The magnetic field was applied perpendicular to the film plane (solid lines) and parallel to it (dotted lines).}
\end{figure}

The crystalline structure was investigated by X-ray diffraction (XRD) using Cu-$K_{\alpha1}$ radiation\,($\lambda=0.15406\,$nm). The symmetrical $2\theta-\omega$ scan is shown in Fig \ref{fig:SQUID XRD}a. The epitaxial relation of the system is given by MgO(001)[100]//Cr(001)[110]//\tMn(001)[110] and according reflexes are expected in the XRD data: Besides the MgO(002) peak of the substrate and the Cr(002) peak of the buffer layer, only~\tMn (001) and~\tMn (002) can be observed, corroborating the epitaxial growth of \tMn~and indicating a predominant (001)-orientation of the film. In addition, the presence of the superlattice peak of (001)~\tMn{} indicates a high degree of chemical ordering.

The magnetic properties were studied via SQUID-VSM at different field orientations and temperatures, as shown in Fig.\ref{fig:SQUID XRD}b. The film is ferromagnetic and does show the expected perpendicular magnetic anisotropy with a saturation magnetisation of $M_\mathrm{s}=272$\,kA/m and a high perpendicular coercive field of $\mu_0H_\mathrm{c}= 1.15$\,T at 100\,K. Similar values were reported for other \tMn~thin films \cite{Hosoda2012,Oshima2020}. At higher temperatures, the saturation magnetisation as well as the coercive field decrease which is expected closer to the Curie temperature of the system \cite{Coey2014}. Compared to bulk samples of MnAl-C \cite{Feng2020}, the film grown here has a much lower saturation magnetisation (cf. film: $M_\mathrm{s} = 272 $\,kA/m, bulk: $M_\mathrm{s} =620$\,kA/m) but a much higher coercivity (cf. film: $\mu_0H_\mathrm{c}=1.15$\,T, extruded bulk, typically $\mu_0H_\mathrm{c}=0.35$\,T).  Please note that the maximum field of $\mu_0H=\pm7$\,T is not sufficient for the saturation of the film when the magnetic field is applied in the magnetically hard film plane.

\section{Anomalous Hall Effect}

\begin{figure}
\centering
\includegraphics{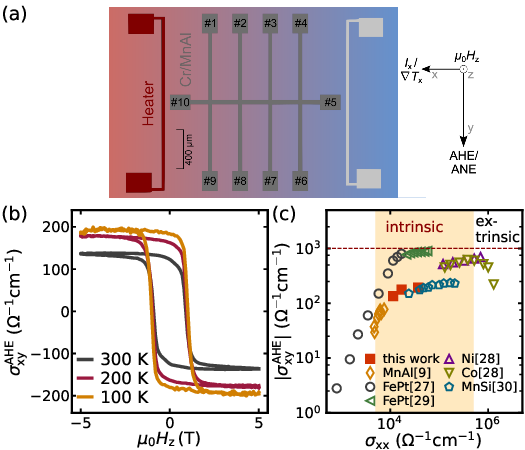}
\caption{\label{fig:structure_AHE}(a) Schematics of the patterned MnAl film and the Pt heater. The contacts of the Hall bar are numbered to enable a simple description of the measurement scheme. (b) Magnetic field dependence of the anomalous Hall conductivity $\sigma_{xy}^\mathrm{AHE}$ for different temperatures. (c) Summary of absolute values of the anomalous Hall conductivity $\sigma_{xy}^\mathrm{AHE}$ as a function of the longitudinal conductivity $\sigma_{xx}$ in various thin film materials \cite{Lu2013,Miyasato2007,He2012,Lee2007,Zhu2016b}, adapted from Oneda \textit{et al.} \cite{Onoda2008}.}
\end{figure}

For the magneto-transport measurements, the \tMn~film was then patterned into Hall bars by a combination of optical lithography and ion beam etching. Subsequently, the on-chip heaters were defined by a lift-off process with \SI{10}{nm} of sputtered Pt. The complete structure is shown in Figure \ref{fig:structure_AHE}a. The magneto-transport response was recorded in a cryostat with variable temperature insert (VTI) to control the sample base temperature. An external magnetic field of up to $\mu_0H_\mathrm{z}=\pm5$\,T was applied perpendicular to the film surface (out-of-plane). The heat gradient for the anomalous Nernst measurements was generated by the on-chip Pt heater. 

First, we evaluate the magneto-transport properties of our \tMn{} thin film by measuring the AHE: A current $I_\mathrm{x}=100$\,$\upmu$A is applied along the Hall bar, i.e.\ between contacts \#5$-$\#8 as shown in Fig. \ref{fig:structure_AHE}a. The transversal voltage $V_{\mathrm{y}}$ (\#4$-$\#6) generated by the Hall effect and the longitudinal voltage $V_{\mathrm{x}}$(\#6$-$\#7) are measured simultaneously as a function of the applied perpendicular magnetic field. We determine the longitudinal resistivity of the sample as $\rho_\mathrm{xx,s}=V_\mathrm{x}wt_\mathrm{s}/I_\mathrm{x}l$ and transversal Hall resistivity as $\rho_\mathrm{xy,s}=V_\mathrm{y}t_\mathrm{s}/I_\mathrm{x}$, where $w$ and $l$ are the width and contact separation of the measured Hall bar, respectively. $t_\mathrm{s}$ denotes the combined thickness of the \tMn~film and the buffer layer. 

It is important to note that the applied current as well as the generated Hall voltage of the \tMn~thin film are affected (shorted) by the highly conductive Cr buffer layer. Therefore, the longitudinal resistivity $\rho_{xx}$ and transversal Hall resistivity $\rho_{xy}$ of the \tMn~thin film are calculated as

\begin{equation}
\label{eq:rho_xx_MnAl}
    \rho_{xx}=t_\mathrm{MnAl}\frac{\rho_{xx,\mathrm{Cr}}\rho_{xx,\mathrm{s}}}{t\rho_{xx,\mathrm{Cr}}-t_\mathrm{Cr}\rho_{xx,\mathrm{s}}}
\end{equation}
\begin{equation}
\label{eq:rho_xy_MnAl}
    \rho_\mathrm{xy}=\rho_\mathrm{xy,s} \frac{(\rho_\mathrm{xx,Cr}t_\mathrm{MnAl}+\rho_\mathrm{xx}t_\mathrm{Cr})^2}{\rho_\mathrm{xx,Cr}^2t_\mathrm{MnAl}t_\mathrm{s}}
\end{equation}

where $\rho_\mathrm{xx,Cr}$ is the longitudinal resistivity of the Cr buffer layer. $\rho_\mathrm{xx,Cr}$ was measured on a separate sample with a single Cr layer. Please note, that we exclude any possible (anomalous) Hall voltages generated within the Cr layer. For a detailed derivation of equations \ref{eq:rho_xx_MnAl} and  \ref{eq:rho_xy_MnAl}, please refer to the supplementary material.
 
The anomalous Hall resistivity of \tMn~$\rho_{xy}^\mathrm{AHE}$ was extracted by subtracting the ordinary $\rho_{xy}^\mathrm{OHE}$ determined by a linear fit to the high field slope of $\rho_{xy}$. Ultimately, we calculate the longitudinal conductivity $\sigma_{{xx}}$ and the anomalous Hall conductivity $\sigma_{xy}^\mathrm{AHE}$ of \tMn~as \cite{Pu2008}

\begin{equation}
\label{eq:sigma_xy}
    \sigma_{xy}^\mathrm{AHE}=-\frac{\rho_{xy}^\mathrm{AHE}}{{\rho_{xy}^\mathrm{AHE}} ^2+\rho_{{xx}}^2}, \sigma_{xx}=1/\rho_{xx}
\end{equation}

The magnetic field dependence of $\sigma_{xy}^\mathrm{AHE}$ at different temperatures is given in Fig. \ref{fig:structure_AHE}b. In accordance to previous studies \cite{Luo2016,Zhu2016b,Takeuchi2022} and to the magnetisation hysteresis of the \tMn~thin film, the anomalous Hall conductivity is scaling with the magnetisation of \tMn~and is showing a hysteresis with a high coercive field of approximately \SI{1}{T}. Consequentially, a decrease of the switching field and a decrease of the magnitude towards higher temperatures can be observed.

The AHE is often discussed in terms of different intrinsic and extrinsic contributions \cite{Nagaosa2010,Onoda2008}. Here, we follow the typical procedure to investigate the origin of the AHE based on the longitudinal and transversal conductivity, in which three distinct regions are distinguished \cite{Onoda2008}. Our \tMn~thin film can be classified into the intermediate regime (see Fig. \ref{fig:structure_AHE}c) according to the conductivity, suggesting that the anomalous Hall effect is dominated by the scattering-independent intrinsic and side-jump contributions \cite{Onoda2008}. This is in accordance to a previous study of \tMn~thin films on GaAs by Zhu \textit{et al.} \cite{Zhu2016b} although our measured $\sigma_{xy}^\mathrm{AHE}$ and longitudinal conductivity $\sigma_{xx}$ are significantly higher. We attribute this to the different preparation procedure of the \tMn~thin films. As already pointed out by  Zhu \textit{et al.} \cite{Zhu2016b}, the intrinsic Hall effect is dependent on the electronic structure of \tMn~ which is easily modified by changes of the chemical composition and chemical disorder \cite{Zhu2016b,Sato2020}.  

\section{Anomalous Nernst Effect}

For the measurement of the ANE, we apply a current of $I_\mathrm{h}=8$\,mA to one of the Pt heaters to generate a heat gradient $\nabla T_{x}$ along the longitudinal direction and measured the transversal voltage $V_{y}$. To determine the heat gradient $\nabla T_{x}$ we adopted methods used previously for ANE measurements \cite{Reichlova2018,Park2020}: two transversal arms of the Hall bar serve as resistive local temperature sensors. The heat gradient is then calculated as $\nabla T_{x}=\Delta T_{x}/L$ where $\Delta T_{x}$ and $L$ are the temperature difference and distance between the two transversal legs. For further details on the thermometry of the ANE measurement, please see the supplementary material.

\begin{figure}
\centering
\includegraphics{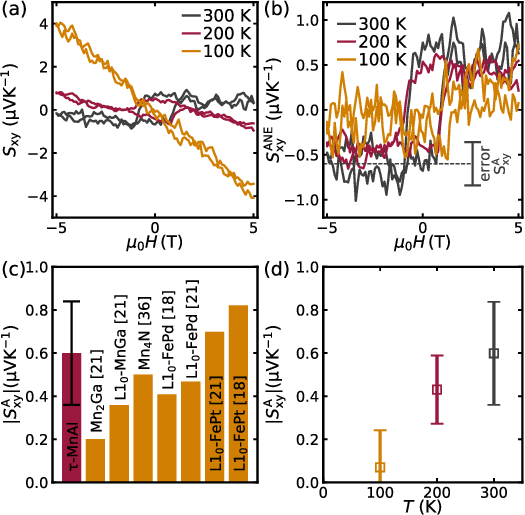}
\caption{\label{fig:Nernst} (a) Magnetic field dependence of $S_{\mathrm{xy}}$ at different temperatures. $V_{\mathrm{y}}$ was measured between \#4 and \#6 at 100\,K and \#3 and \#7 at 200\,K and 300\,K. (b) Magnetic field dependence of $S_{\mathrm{xy}}^{\mathrm{ANE}}$ for different temperatures. The error of $S_{\mathrm{xy}}^{\mathrm{A}}$ was calculated based on the error of the heat gradient and the noise of the measured Nernst voltage and is exemplary given for 300\,K. (c) Comparison of the absolute Anomalous Nernst coefficient $|S_{\mathrm{xy}}^{\mathrm{A}}|$ for different PMA materials and the results of this study, for a sample temperature of 300\,K \cite{Hasegawa2015,Shi2020,Isogami2021}. (d) Evolution of the anomalous Nernst coefficient with temperature.}
\end{figure}

Similar to the Hall effect, the electrically conductive Cr buffer layer also impacts the magnitude of the anomalous Nernst effect. Thus, the Nernst effect signal is calculated as

\begin{equation}
\label{eq:Nernst1}
    S_{{xy}} = \frac{V_{y}(\rho_{xx,\mathrm{Cr}}t_\mathrm{MnAl}+\rho_{xx}t_\mathrm{Cr})}{\rho_{xx,\mathrm{Cr}}t_\mathrm{MnAl}}\frac{1}{\nabla T_{x}d}\\
\end{equation}

where $d=1400$\,$\upmu\mathrm{m}$ is the contact distance of the transversal contacts. A derivation of equation \ref{eq:Nernst1} is given in the supplementary material. 

The field-dependent Nernst signal $S_{\mathrm{xy}}$ for different temperatures thus obtained is given in Fig. \ref{fig:Nernst}a. $S_{xy}$ consists of a negative linear function superimposed by a hysteresis at 200\,K and 300\,K. No hysteresis can be resolved at a sample temperature of 100\,K. Similar to the Hall effect, the Nernst effect can phenomenologically be understood as the combination of the ordinary Nernst contribution $S_{{xy}}^{\mathrm{ONE}}$ scaling linear with the external magnetic field $H$ and the magnetisation dependent $S_{{xy}}^{\mathrm{ANE}}$ \cite{Hasegawa2015}

\begin{equation}
\label{eq:Nernst2}
    S_{{xy}} = S_{{xy}}^{\mathrm{ONE}} + S_{{xy}}^{\mathrm{ANE}} = \mu_0HN_\mathrm{ONE} + S_{xy}^{\mathrm{A}} \frac{M}{M_\mathrm{s}}
\end{equation}

where $N_\mathrm{ONE}$ and $S_{xy}^{\mathrm{A}}$ are the ordinary and anomalous Nernst coefficients, respectively. The hysteretic contribution contribution to $S_{xy}$ thus corresponds to the ANE where as the general negative slope in each measurement can be assigned to the ONE. 

 \begin{table*}[]
 \caption{Measured transport coefficients of the (isolated) \tMn~thin film at different temperatures. (*) These values are not measured but estimated by using the Mott relation \cite{Pu2008}.}
    \centering
    \begin{tabular}{c@{\hskip 0.2in}c@{\hskip 0.2in}c@{\hskip 0.2in}c@{\hskip 0.2in}c@{\hskip 0.2in}c}
    \hline
    \hline
         $T$ (K) & $\sigma_{xx}$ ($\Omega^{-1}$cm$^{-1})\times 10^3$ & $\sigma_{xy} (\Omega^{-1}$cm$^{-1})$ &$S_{xx} (\upmu$V/K)   & $S_{xy}^\mathrm{A} (\upmu$V/K) &  $\alpha_{xy} $(A/Km)\\
         \hline
         100& 37.174 & 193.05 & 8.01  & 0.1 $\pm$ 0.17 * & 0.155*\\
         200& 17.088 & 176.69 & 9.64  & 0.43 $\pm$ 0.17 & 0.904\\
         300& 11.494 & 134.61 & 22.84 & 0.6  $\pm$ 0.24 & 0.998\\
         \hline
         \hline
    \end{tabular}
    \label{tab:results}
\end{table*}

The anomalous Nernst contribution $S_{{xy}}^{\mathrm{ANE}}$ is calculated by substracting the ONE from the Nernst effect and is given in Fig. \ref{fig:Nernst}b as a function of the external field. $S_{{xy}}^{\mathrm{ANE}}$ switches in sign at around $\pm1$\,T, in close analogy to the magnetization curve and the AHE. This supports the notion that the thermal voltage is generated by the ANE of the \tMn~thin film. For a sample temperature of 100\,K, no clear ANE signal can be observed.

From the data in Fig. 3b, we obtain an anomalous Nernst coefficient $S_{{xy}}^{\mathrm{A}}=(0.60\pm 0.24)\,\upmu$V/K at 300\,K and $(0.43\pm 0.16)\,\upmu$V/K at 200\,K. For the error calculation we consider an error of the heat gradient of $\pm0.1\mathrm{Kmm}^{-1}$ as well as the noise level of the measured anomalous Nernst voltage which yields a relative high error of $S_{{xy}}^{\mathrm{A}}$. Given this high error, we expect the
ANE at 100 K to be below the resolution limit of our ANE measurement as visible in Fig.~\ref{fig:Nernst}b.

Comparing the ANE at 200\,K and 300\,K, we observe a decrease of the ANE towards lower temperatures, which is in contrast to the trend observed in AHE and magnetometry (see Fig. \ref{fig:Nernst}d). Similar correlations are reported for other ferromagnetic materials at temperatures well below $T_\mathrm{C}$ \cite{Lee2004,Hasegawa2015,Sakai2020,Park2020,Ghosh2019}, including L1$_0$-ordered alloys \cite{Hasegawa2015}. There, the ANE vanishes towards T=0\,K as the entropy change of the magnet reaches zero \cite{Hasegawa2015}.

To gain further insights into the temperature dependence of the ANE, we also measured the Seebeck effect of the \tMn~thin film. For these experiments, a heating current of 8\,mA was applied and the voltage $V_{x}$ was measured between contact \#2 and \#4 (see Fig. \ref{fig:structure_AHE}a). Again, the electrical conductive Cr layer impacts the magnitude of the Seebeck effect and the Seebeck coefficient $S_{xx}$ is thereby calculated by 
\begin{equation}
\label{eq:S_x_f}
    S_{xx}=-\frac{V_{x}(\rho_{xx}t_\mathrm{Cr}+\rho_{xx,\mathrm{Cr}}t_\mathrm{MnAl})-V_\mathrm{x,Cr}\rho_{xx}t_\mathrm{Cr}}{\rho_{xx\mathrm{,Cr}}t_\mathrm{MnAl}\Delta T_x}
\end{equation}

where $V_{x,\mathrm{Cr}}$ is the voltage generated by the Seebeck effect of the Cr layer \cite{Arajs1972}. More details on the Seebeck measurement as well as a derivation of Eq. \ref{eq:S_x_f} are given in the supplementary material. Using $S_{xx}$, we can calculate the anomalous Nernst conductivity $\alpha_{xy}$ for the temperatures of 200\,K and 300\,K, given by the general relation of the transport coefficients \cite{Pu2008}:

\begin{equation}
\label{eq:alpha}
\alpha_{xy}=S_{xx}\sigma_{xy}+S_{xy}\sigma_{xx}
\end{equation}

All transport coefficients are summarised in Table \ref{tab:results}.

By fitting of both the measured $S_{xy}$ and $\alpha_{xy}$ at 200\,K and 300\,K to the Mott relation \cite{Pu2008}:

\begin{equation}
    S_\mathrm{xy}=\frac{\rho_\mathrm{xy}}{\rho_\mathrm{xx}}\left(\frac{\pi^2 k^2_\mathrm{B}}{3e}\frac{\lambda'}{\lambda}-(n-1)S_\mathrm{xx}\right)
\end{equation}
\begin{equation}
    \alpha_\mathrm{xy}=\frac{\rho_\mathrm{xy}}{\rho_\mathrm{xx}^2}\left(\frac{\pi^2 k^2_\mathrm{B}}{3e}\frac{\lambda'}{\lambda}-(n-2)S_\mathrm{xx}\right)
\end{equation}

where $k_\mathrm{B}$ and $e$ denote the Boltzmann constant and the elementary charge, we can determine a pair of $\frac{\lambda'}{\lambda}= 6.43 \times 10^{-19}$\,J$^{-1}$ and $n=2$ for the \tMn~thin film. Given these values, we can finally estimate the anomalous Nernst coefficient at 100\,K to be $S_{{xy}}^{\mathrm{A}}=0.1 \pm 0.17 \,\upmu$V/K. The high error exceeds the estimated value of $S_{{xy}}^{\mathrm{A}}$ and hence explains why the ANE is not resolvable at 100\,K.

The magnitudes of $S_{{xy}}^{\mathrm{A}}$ are well in the range reported for other ferromagnetic thin film materials \cite{Hasegawa2015,Isogami2021,Ikhlas2017,Shi2020} as compiled in Fig. \ref{fig:Nernst}c. As shown for other L1$_0$-ordered thin films, the magnitude of $S_{{xy}}^{\mathrm{A}}$ is scaling with the magnetisation and magnetic anisotropy of the material \cite{Ikhlas2017,Hasegawa2015,Shi2020}. However, as $S_{{xy}}^{\mathrm{A}}$ is strongly dependent on film thickness \cite{Chuang2017,Kannan2017}, a direct comparison of different materials is challenging. In addition, the magnetic properties of L1$_0$-ordered materials are highly dependent on the chemical composition and structural ordering \cite{Hosoda2012,Oogane2017,Zhou2021,Shi2020}. The variance of the AHE in \tMn~which we discussed earlier can also be extended to the ANE, given the similar physical origin of both effects. Therefore further anomalous Nernst measurement in dependence of film thickness, chemical composition and structural disorder will be important to gain a better understanding of the magneto-thermal transport properties of \tMn. 

\section{\label{Conclusion}Conclusion}

We prepared \tMn~thin films with perpendicular magnetic anisotropy via sputter deposition, and measured the anomalous Nernst effect at various temperatures.  The ANE features a robust hysteresis and spontaneous signal in zero magnetic field, in close qualitative agreement with anomalous Hall effect and SQUID magnetometry data.  At 300 K the anomalous Nernst coefficient reaches $S_{{xy}}^{\mathrm{A}}=(0.60\pm 0.24)\,\upmu$V/K, a value well within the range of ANE coefficients reported in other  PMA materials with similar magnetic properties. Our results show that \tMn~in thin film form is an interesting material for spin-caloritronic research and devices.

\section{\label{Acknowlegments}Acknowledgements}

We thank Torsten Mix for helpful discussions. HR acknowledges Czech Science Foundation GACR 22-17899K. This work was funded by the Deutsche Forschungsgemeinschaft (DFG, German Research Foundation) – Project number: 380033763.

\bibliography{ANE_MnAl.bbl}
\end{document}


\title{Anomalous Nernst effect in perpendicularly magnetised $\uptau$-MnAl thin films - Supplementary Material}

\author{D. Scheffler}
\affiliation{Institute for Solid State and Materials Physics, Technische Universität Dresden, 01062 Dresden, Germany}

\author{S.Beckert}%
\affiliation{Institute for Solid State and Materials Physics, Technische Universität Dresden, 01062 Dresden, Germany}

\author{H.Reichlova}%
\affiliation{Institute for Solid State and Materials Physics, Technische Universität Dresden, 01062 Dresden, Germany}
\affiliation{Institute of Physics ASCR, v.v.i., Cukrovarnick\'a 10, 162 53, Praha 6, Czech Republic}

\author{T.G. Woodcock}
\affiliation{Leibniz Institute for Solid State and Materials Research Dresden (IFW Dresden), 01069 Dresden, Germany}

\author{S.T.B. Goennenwein}
\affiliation{Department of Physics, University of Konstanz, 78457 Konstanz, Germany}

\author{A. Thomas}%
\affiliation{Institute for Solid State and Materials Physics, Technische Universität Dresden, 01062 Dresden, Germany}
\affiliation{Leibniz Institute for Solid State and Materials Research Dresden (IFW Dresden), 01069 Dresden, Germany}

\date{\today}

\maketitle

\section{Derivation of Anomalous Hall and Nernst Effect considering a conductive buffer layer}
    
In general, the longitudinal resistivity $\rho_{xx}$, Hall resistivity $\rho_\mathrm{xy}$, Hall conductivity $\sigma_{xy}$ and Nernst effect $S_{xy}$ of a film stack are caluculated as:

\begin{equation}
\label{eq:rho_xx}
    \rho_{xx}=\frac{V_{x}w t}{I_{x}l}
\end{equation}

\begin{equation}
\label{eq:rho_xy}
    \rho_{xy}=\frac{V_{y}t}{I_{x}}
\end{equation}

\begin{equation}
\label{eq:sigma_xy}
    \sigma_{xy}=-\frac{\rho_{xy}}{\rho_{xy}^2+\rho_{xx}^2}
\end{equation}

\begin{equation}
\label{eq:S_xy}
    S_{xy}=\frac{V_{y}}{\nabla T_{x} d}
\end{equation}

where $w$, $l$, $d$ and $t$ are the bar width, the longitudinal contact separation, transversal contact separation and film thickness of the Hall bar, respectively. $I_{x}$ is the applied current along the x-direction, $\nabla T_\mathrm{x}$ the temperature gradient along the ${x}$-direction. The voltage is recorded in ${x}$-direction ($V_{x}$) and ${y}$-direction ($V_{y}$). 
However, if the film consist of a ferromagnetic film layer with thickness $t_\mathrm{F}$ and a buffer layer with thickness $t_\mathrm{B}$, the applied current and the transversal Hall/Nernst voltage of the ferromagnet are shorted by the buffer layer.

\subsection{Influence on longitudinal resistivity.}

\begin{figure}[h]
\begin{center}
\begin{circuitikz}
\draw
(-1,0)          to[ioosource,l=$I_\mathrm{x}$]            (4,0)
(4,0)           --                                          (4,-2)
(4,-2)          --                                          (2.5,-2)
(2.5,-2)        --                                          (2.5,-1.5)
(2.5,-2)        --                                          (2.5,-2.5)
(0.5,-1.5)        to[european resistor,l=$R_\mathrm{B}$]      (2.5,-1.5)
(0.5,-2.5)        to[european resistor,l=$R_\mathrm{F}$]      (2.5,-2.5)
(0.5,-1.5)        --                                          (0.5,-2.5)
(0.5,-2)        --                                          (-1,-2)
(-1,-2)         --                                          (-1,0)
(3.25,-2)       --                                          (3.25,-3.75)
(-0.25,-3.75)     to[rmeter, t=V,l=$V_\mathrm{x}$]           (3.25,-3.75)
(-0.25,-2)      --                                          (-0.25,-3.75)
;
\end{circuitikz}
\end{center}
\caption{\label{cir:resistivity}
Electrical circuit for the longitudinal direction measurement}
\end{figure}
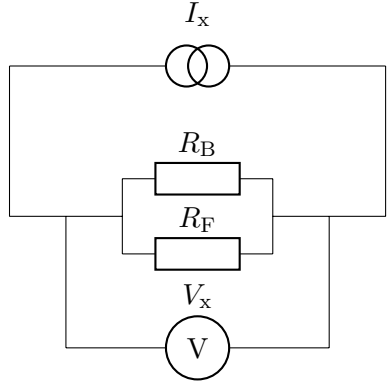

In the longitudinal direction, the conductive buffer layer carries a part of the current. We model this by two individual resistors representing the buffer layer ($R_\mathrm{B}$) and the ferromagnetic film layer ($R_\mathrm{F}$) which are connected parallel (see electrical circuit Fig. \ref{cir:resistivity}). Then the total resistance/resistivity $R$ is defined by:

\begin{equation}
\label{eq:R_parallel}
    \frac{1}{R}=\frac{1}{R_\mathrm{F}}+\frac{1}{R_\mathrm{B}}
\end{equation}
    
\begin{equation}
\label{eq:rho_parallel}
    \frac{t}{\rho_{xx}}=\frac{t_\mathrm{F}}{\rho_{xx,\mathrm{F}}}+\frac{t_\mathrm{B}}{\rho_{xx,\mathrm{B}}}
\end{equation}
Then, the resistivity of the ferromagnetic film $\rho_\mathrm{xx,F}$ is calculated as:

\begin{equation}
\label{eq:rho_xx_f}
    \rho_{xx,\mathrm{F}}=\frac{t_\mathrm{F}}{\frac{t}{\rho_{xx}}-\frac{t_\mathrm{B}}{\rho_{xx,\mathrm{B}}}}=t_\mathrm{F}\frac{\rho_{xx}\rho_{xx,\mathrm{B}}}{t\rho_{xx,\mathrm{B}}-t_\mathrm{B}\rho_{xx}}
\end{equation}

\section{Influence on Hall/Nernst effect}

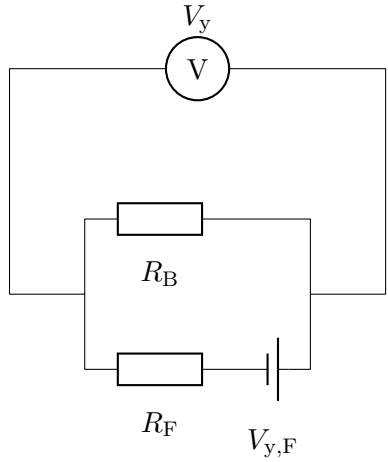
\begin{figure}[ht]
\begin{center}
\begin{circuitikz}
\draw
(-1,1)          to[rmeter, t=V,l=$V_\mathrm{y}$]          (4,1)
(4,1)           --                                          (4,-2)
(4,-2)          --                                          (3,-2)
(3,-2)          --                                          (3,-1)
(3,-2)          --                                          (3,-3)
(3,-3)          to[battery1,l=$V_\mathrm{y,F}$]         (2,-3)
(3,-1)          --                                          (2,-1)
(2,-1)          to[european resistor,l=$R_\mathrm{B}$]      (0,-1)
(2,-3)          to[european resistor,l=$R_\mathrm{F}$]      (0,-3)
(0,-1)          --                                          (0,-3)
(0,-2)          --                                          (-1,-2)
(-1,-2)         --                                          (-1,1)
;
\end{circuitikz}
\end{center}
\caption{\label{cir:Hall}
Electrical circuit for the anomalous Hall/Nernst effect}
\end{figure}

Here we assume an electrical circuit as in Fig.\ref{cir:Hall}. The voltage generated by the Hall or Nernst effect of the functional film layer ($V_{y,\mathrm{F}}$) is treated as a battery which has an internal resistance equal to the resistance of this layer ($R_\mathrm{F}$), which is shorted by the parallel buffer layer resistance $R_\mathrm{B}$.

$V_\mathrm{y}^\mathrm{F}$ in this electrical circuit is thereby calculated by:

\begin{equation}
    V_{y,\mathrm{F}}=V_\mathrm{y} (\frac{R_\mathrm{F}+R_\mathrm{B}}{R_\mathrm{B}})
\end{equation}

\begin{equation}
\label{eq:V_y_f}
    V_{y,\mathrm{F}}=V_{y} \frac{(\rho_{xx,\mathrm{B}}t_\mathrm{F}+\rho_{xx,\mathrm{F}}t_\mathrm{B})}{\rho_{xx,\mathrm{B}}t_\mathrm{F}}
\end{equation}

For the anomalous Hall effect, using this equation and equation \ref{eq:rho_xy} for the ferromagnetic thin film yields:

\begin{equation}
\label{eq:rho_y_f}
    \rho_{xy,\mathrm{F}}=V_{y} \frac{(\rho_{xx,\mathrm{B}}t_\mathrm{F}+\rho_{xx,\mathrm{F}}t_\mathrm{B})}{\rho_{xx,\mathrm{B}}I_{x,\mathrm{F}}}
\end{equation}

$I_{x,\mathrm{F}}$ is the current flowing through the ferromagnetic film and is not equal to the applied current $I_{x}$. It can be calculated assuming the parallel resistor electrical circuit in Fig. \ref{cir:resistivity}:

\begin{equation}
\label{eq:I_x_F}
    I_{x,\mathrm{F}}=\frac{I_{xx}R_\mathrm{F}}{R_\mathrm{B}+R_\mathrm{F}}=I_{x}\frac{\rho_{xx,\mathrm{B}}t_\mathrm{F}}{\rho_{xx,\mathrm{B}}t_\mathrm{F}+\rho_{xx,\mathrm{F}}t_\mathrm{B}}
\end{equation}

then using \ref{eq:rho_y_f} and \ref{eq:I_x_F}:

\begin{equation}
\label{eq:rho_y_f2}
    \rho_{xy,\mathrm{F}}=V_{y} \frac{(\rho_{xx,\mathrm{B}}t_\mathrm{F}+\rho_{xx,\mathrm{F}}t_\mathrm{B})^2}{I_{x}\rho_{xx,\mathrm{B}}^2t_\mathrm{F}}=\rho_{xy} \frac{(\rho_{xx,\mathrm{B}}t_\mathrm{F}+\rho_{xx,\mathrm{F}}t_\mathrm{B})^2}{\rho_{xx,\mathrm{B}}^2t_\mathrm{F}t}
\end{equation}
\\
\\
For the anomalous Nernst coefficient, we assume a constant heat gradient in both buffer and ferromagnetic thin film layer, hence the anomalous Nernst coefficient is calculated using equation \ref{eq:S_xy} and \ref{eq:V_y_f}: 

\begin{equation}
    S_{xy,\mathrm{F}}=\frac{V_{y,\mathrm{F}}}{\nabla T_{x} d}=\frac{V_{y}(\rho_{xx,\mathrm{B}}t_\mathrm{F}+\rho_{xx,\mathrm{F}}t_\mathrm{B})}{\rho_{xx,\mathrm{B}}t_\mathrm{F}}\frac{1}{\nabla T_{x}d}
\end{equation}

\section{Thermometry for Anomalous Nernst measurement}

For the measurement of the heat gradient generated by the on-chip Platinum heaters, a current of 50\,$\upmu$A was applied longitudinal between two  contacts of the hall bar, i.e. contact \#2 and \#4. The voltage was measured between contacts \#1$-$\#2 and \#4$-$\#5. The two measured voltages are representative for the transversal contact legs at contact \#2 and \#4. For calibration as resistive temperature sensors, the voltages where monitored as a function of the base temperature of the cryostat, shown in Fig. \ref{fig:temp_calib}. The curve of each sensor was fitted to a polynomial function which was utilised as calibration function for the temperature. 

\begin{figure}[ht]
\includegraphics{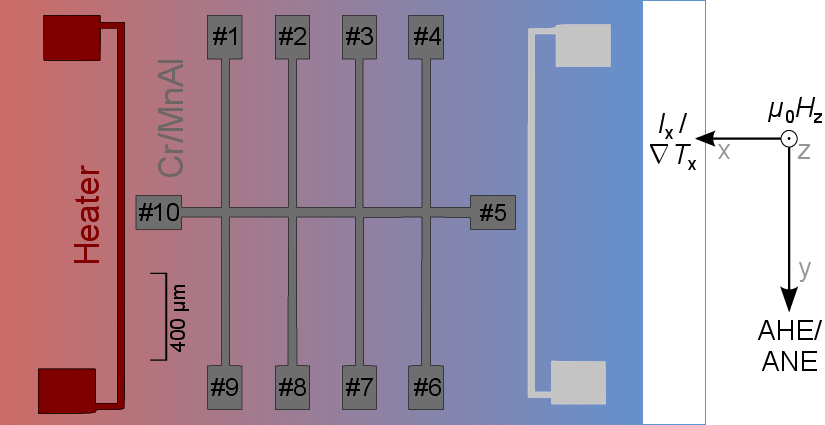}
\caption{\label{fig:structure_general}Schematics of the structured MnAl thin film. For better comprehension, all contacts of the Hall bar are numbered.}
\end{figure}

\begin{figure}[ht]
    \begin{subfigure}[b]{0.49\textwidth}
        \centering
        \includegraphics[width=\textwidth]{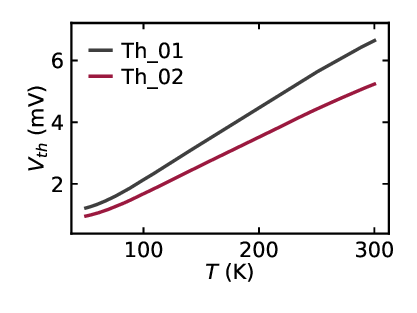}
        \end{subfigure}
    \begin{subfigure}[b]{0.49\textwidth}
        \centering
        \includegraphics[width=\textwidth]{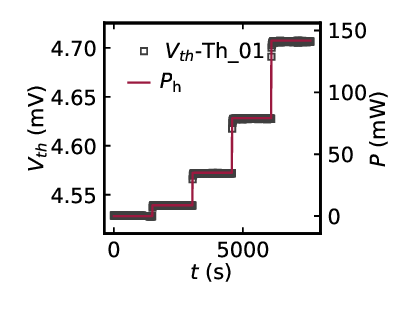}
    \end{subfigure}
    \caption{(left) Voltage as a function of the cryostat base temperature (right) Voltage measured for the stepwise increase of the heating current at 200\,K.}
    \label{fig:temp_calib}

\end{figure}

\begin{figure}[h]
\includegraphics{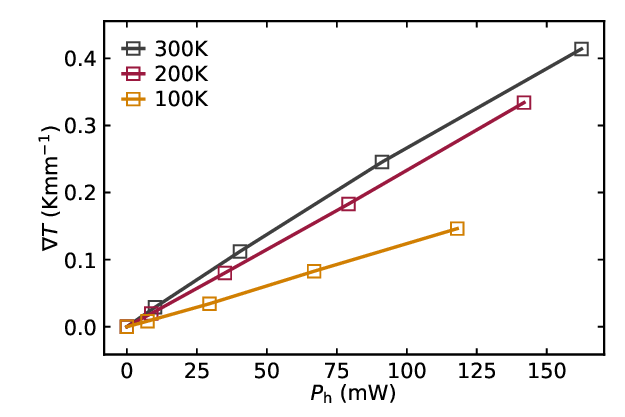}
\caption{\label{fig:temp_calib}Temperature gradient as a function of the heating power for different cryostat base temperatures.}
\end{figure}

Then, the base temperature of the cryostat was set to a constant temperature and the heating current was increased in 2\,mA steps up to $I_\mathrm{h}=8$\,mA, as shown in Fig. \ref{fig:temp_calib} exemplary for $T=200$\,K. For each step, the temperature of both sensors was calculated based on the calibration function. The temperature gradient was then calculated as $\nabla T_{x}=\Delta T_{x}/L$ where $\Delta T_{x}$ and $L$ are the temperature difference and distance between the two measured contact pairs. The measured temperature gradient as a function of the heating power is shown in Fig. \ref{fig:temp_calib}. The temperature gradient is increasing linearly with the applied heating power supporting that the presented measurement routine is suitable for the calculation of the temperature gradient. For the maximum heating current of $I_\mathrm{h}=8$\,mA which was also used for the ANE measurements, the temperature gradient was measured as $\nabla T_{x}=(0.15;0.33;0.41)$\,$\mathrm{Kmm}^{-1}$ for the temperatures of $T=(100;200;300)$\,K, respectively. 

However, it must be noted that the voltages were measured in three point configuration. Therefore we assume a high error of the measured temperature gradient of $\pm0.1\mathrm{Kmm}^{-1}$.

\newpage
\section{Seebeck effect measurement and influence of Chromium buffer}

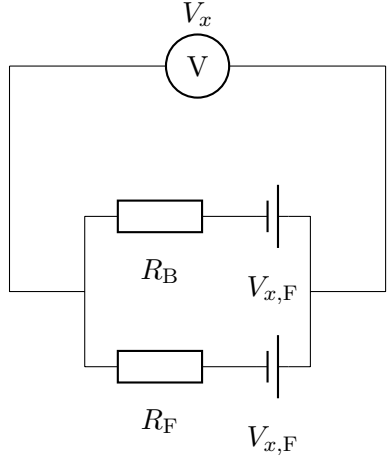
\begin{figure}[ht]
\begin{center}
\begin{circuitikz}
\draw
(-1,1)          to[rmeter, t=V,l=$V_{x}$]          (4,1)
(4,1)           --                                          (4,-2)
(4,-2)          --                                          (3,-2)
(3,-2)          --                                          (3,-1)
(3,-2)          --                                          (3,-3)
(3,-1)          to[battery1,l=$V_{x,\mathrm{F}}$]         (2,-1)
(3,-3)          to[battery1,l=$V_{x,\mathrm{F}}$]         (2,-3)
(2,-1)          to[european resistor,l=$R_\mathrm{B}$]      (0,-1)
(2,-3)          to[european resistor,l=$R_\mathrm{F}$]      (0,-3)
(0,-1)          --                                          (0,-3)
(0,-2)          --                                          (-1,-2)
(-1,-2)         --                                          (-1,1)
;
\end{circuitikz}
\end{center}
\caption{\label{cir:Seebeck}
Electrical circuit for the Seebeck effect.}
\end{figure}

In the Seebeck measurement, the measured voltage is generated by both the Seebeck effect of the $\tau$-MnAl and the buffer layer. For that reason, we assume an electrical circuit as in Fig.\ref{cir:Seebeck}. The voltage generated by the Seebeck effect of the functional film layer ($V_{x,\mathrm{F}}$) and the buffer layer ($V_{x,\mathrm{B}}$) are treated as a battery which have an internal resistance equal to the resistance of layers ($R_\mathrm{F}$,$R_\mathrm{B}$) which are then connected in parallel.

$V_{x}$ is thereby calculated as:

\begin{equation}
    V_{x}=V_{x,\mathrm{F}} (\frac{R_\mathrm{B}}{R_\mathrm{B}+R_\mathrm{F}})+V_{x,\mathrm{B}} (\frac{R_\mathrm{F}}{R_\mathrm{B}+R_\mathrm{F}})
\end{equation}
 and hence $V_{y,\mathrm{F}}$ is calculated as:

\begin{equation}
\label{eq:V_x_f}
    V_{x,\mathrm{F}}=\frac{V_{x}(\rho_{xx,\mathrm{F}}t_\mathrm{B}+\rho_{xx,\mathrm{B}}t_\mathrm{F})-V_{x,\mathrm{B}}\rho_{xx,\mathrm{F}}t_\mathrm{B}}{\rho_{xx,\mathrm{B}}t_\mathrm{F}}
\end{equation}

Ultimately the Seebeck coefficient is calculated by

\begin{equation}
\label{eq:S_xx_F}
    S_{xx,\mathrm{F}}=-\frac{V_{x,\mathrm{F}}}{\Delta T}
\end{equation}

The measured voltages for the different base temperatures as well as the calculated Seebeck coefficient are summarised in table \ref{tab:Seebeck}. The temperature difference $\Delta T$ was measured as described in the previous section. The Seebeck coefficient of Cr was taken from reference\cite{Arajs1972}.

\begin{table}[ht]
    \centering
    \begin{tabular}{c|cccc}
    \hline
    \hline
        $T$ (K) & $V_{x}$ ($\upmu \mathrm{V}$) &  $V_{x,\mathrm{MnAl}}$ ($\upmu \mathrm{V})$ & $S_{xx,\mathrm{Cr}}$ ($\upmu \mathrm{V} \mathrm{K}^{-1}$) & $S_{xx,\mathrm{MnAl}}$ ($\upmu \mathrm{V} \mathrm{K}^{-1}$) \\
        \hline
        100 & -0.587 & -0.718 & 5     & 8.015\\
        200 & -2.516 & -2.187 & 12.5  & 9.64\\
        300 & -7.012 & -7.147 & 22    & 22.84\\
    \hline
    \hline
    \end{tabular}
    \caption{Results of the Seebeck effect measurement}
    \label{tab:Seebeck}
\end{table}